# Enhancing Mean-Reverting Time Series Prediction with Gaussian Processes: Functional and Augmented Data Structures in Financial Forecasting

**Narayan Tondapu**
narayan.tondapu@gmail.com
Microsoft, Redmond, Washington, USA

**Abstract**

In this paper, we explore the application of Gaussian Processes (GPs) for predicting mean-reverting time series with an underlying structure, using relatively unexplored functional and augmented data structures. While many conventional forecasting methods concentrate on the short-term dynamics of time series data, GPs offer the potential to forecast not just the average prediction but the entire probability distribution over a future trajectory. This is particularly beneficial in financial contexts, where accurate predictions alone may not suffice if incorrect volatility assessments lead to capital losses. Moreover, in trade selection, GPs allow for the forecasting of multiple Sharpe ratios adjusted for transaction costs, aiding in decision-making. The functional data representation utilized in this study enables longer-term predictions by leveraging information from previous years, even as the forecast moves away from the current year's training data. Additionally, the augmented representation enriches the training set by incorporating multiple targets for future points in time, facilitating long-term predictions. Our implementation closely aligns with the methodology outlined in [1], which assessed effectiveness on commodity futures. However, our testing methodology differs. Instead of real data, we employ simulated data with similar characteristics. We construct a testing environment to evaluate both data representations and models under conditions of increasing noise, fat tails, and inappropriate kernels—conditions commonly encountered in practice. By simulating data, we can compare our forecast distribution over time against a full simulation of the actual distribution of our test set, thereby reducing the inherent uncertainty in testing time series models on real data. We enable feature prediction through augmentation and employ sub-sampling to ensure the feasibility of GPs. The experiments demonstrate the effectiveness of the functional and augmented data representations, quantify the impact of noise and fat tails on these models, and identify scenarios where simpler models suffice. We explore the consequences of choosing an incorrect initial kernel and illustrate how functional augmentation can mitigate this issue under certain circumstances. Furthermore, we showcase how augmentation enhances predictive capability in scenarios with limited training data and present innovative applications of augmented GP in trading exchange-traded futures.

## 1. Introduction

This study aims to assess the effectiveness of Gaussian Processes (GPs) in predicting both simulated financial data and real-life financial trends. The approach follows the methodology outlined by [1], who experimented with unconventional data representations as inputs to GPs for forecasting mean reverting commodity futures and equities, respectively. [1] observed favorable outcomes in terms of test statistics and trading performance when applied to actual commodity futures. However, in real-time financial data, we encounter the challenge of dealing with a single realization of a noisy process. Unlike in simulations, where we have control over parameters; whereas in real-life scenarios, we lack the luxury of revisiting past data [2]–[6]. Confidence in our predictions grows with larger datasets and a stable underlying process, yet uncertainties persist, especially in the finance domain, where the effectiveness of back tests may not always translate into real-world success. Inspired by the potential for long-term forecasting demonstrated in [1], [7]–[11], we examine the suitability of these data representations and models using simulated data that mimics the characteristics of real financial data. By manipulating test set parameters and generating the full distribution of the test set through simulation, we can verify whether the GP accurately captures the variability in the noisy time-series. Subsequently, we transition to real-life financial time series to demonstrate the practical application of GPs in trading strategies. We highlight the advantages of incorporating mean function forecasts and quantifying forecast uncertainty, illustrating how these insights can inform trade decision-making processes directly.

In many cases, time series models limit us to predicting outcomes within a set timeframe. While they may capture short-term fluctuations well, their predictions often deteriorate into simply reflecting the long-term average of the process. Other regression methods, commonly utilized in financial contexts, hinge on predicting outcomes based on certain conditions: "What will the value of my target 'y' be, given that my features have specific values 'x'?". Therefore, this study explores the application of GPs regression to predict entire trajectories into the future. Additionally, it investigates scenarios where this prediction outperforms conventional autoregressive models or standard one-dimensional GPs, leveraging functional and augmented data representations as discussed by [1]. Utilizing an implementation of GPs along with diverse data representations and a simulated time series, we train various models on this data and endeavor to forecast future distributions. Because our data is simulated, we have the luxury of rerunning our time series from a specific point multiple times to generate the complete distribution of the test set. This furnishes us with both predictive and test distributions for each individual training example, allowing us to assess the model's predictive capacity by examining Mean Squared Error (MSE) and standard deviation predictions for several future time points.

However, we commence with a structured time series and subsequently introduce an Ornstein Uhlenbeck (OU) process into this framework. Since, the study documented in [1] posited that under the functional representation, information from prior years could be leveraged as the GPs moves away from the training data. By imparting structure to the data, we can evaluate this claim. However, challenges escalate as the noise in the OU process is heightened. The initial segment of the study assesses how these methodologies compare to each other and to the conventional mean-reverting auto-regressive models typically utilized in finance, particularly in scenarios with escalating noise levels. Subsequently, the investigation delves deeper into the challenges faced by GPs in more realistic settings, characterized by heavier-tailed non-Gaussian noise and the risk of selecting an unsuitable kernel prior. Once again, we evaluate the performance of these intricate models in relation to each other and against an auto-regressive model. Moving forward, a brief investigation is conducted on the ramifications of limited training data on the models, showcasing the potential benefits of augmentation techniques. Finally, the study shifts focus to a practical issue of utilizing GPs for trading Exchange-Traded Funds (ETFs), demonstrating how incorporating not only a prediction of the mean function but also the associated uncertainty into a trading strategy can be directly implemented.

The following is the paper. We shall discuss the study's backdrop in the following section. Data analysis and the material and techniques portion are described in Section 3. The experimental analysis is carried out in Section 4. A discussion part is included in Section 5, and the study is concluded with some conclusions and ideas for future work in Section 6.

## 2. Background

In this section, we outline the basics of GPs, including their kernels, training procedures, and optimization techniques. Additionally, we introduce automatic relevance determination, a method akin to $l_1$ regularization or the Lasso for traditional regression problems. Various presentations of these concepts are available in studies such as [12], [13]. When dealing with continuous value predictions from observed data, we enter a regression setting. One of the simplest forms of regression is linear regression. A regression typically comprises two components: systematic variation, denoted as $f(x)$, and random variation, reflecting the inherent unpredictability of our system. Here, we briefly introduce linear regression, a fundamental concept, to lay the groundwork for GPs. Consider a training dataset with '$n$' samples, where each observation '$x$' corresponds to an output '$y$'. A regression function '$f(x)$' is considered linear if it can be expressed as '$f(x) = w'x$', with the target values subject to Gaussian noise, represented in Equation (1).

$$y = f(x) + \varepsilon', where\ \varepsilon \sim N(0, \sigma^2) \tag{1}$$

Given '$w$' and assuming independence, the likelihood can be written as the minimization of the sum of squared errors. Taking the partial derivative with respect to '$w$' and setting it to zero provides the solution for '$w$'. It's important to recognize that the solution for '$w$' yields a single point estimate for the model parameters. However, in reality, different samples of training data could lead to slightly different estimates. Furthermore, if prior beliefs about '$w$' exist, they are not reflected in this solution. However, Bayesian methodology extends the above approach by considering a distribution over the parameters, offering a more comprehensive view. With training data '$y$' and '$x$', we can derive the distribution of weights using Bayes' rule. Similarly, by introducing a Gaussian prior to '$w$', the distribution of weights takes a specific form. This distribution allows us to obtain noiseless and noisy predictions, providing insights into the uncertainty associated with our predictions [14]–[19]. A notable restriction of our current method arises from its linear nature. This means we assume there's a straight-line relationship between what we observe and what we're trying to predict. However, we can broaden the possibilities of our prediction by introducing some non-straight-line functions. Specifically, we can express our prediction formula as adding up various distances between our data points and some reference points, then multiplying them by weights. In simpler terms, our prediction now follows a curvy path, even though the calculations themselves remain straightforward. When we train our model with data, we express our predictions as a combination of these calculated distances, with some room for error. This approach might sound complex, but it simplifies the predictions into manageable parameters. Whereas, in statistical modeling, we delve into two main approaches: maximum likelihood and Bayesian methods [20]–[23]. Both involve similar calculations but differ in their underlying assumptions. Bayesian reasoning starts with setting initial beliefs about weights, often assumed to follow a bell-shaped curve, and updates these beliefs using observed data to refine predictions. Technically, our predictions follow a specific distribution pattern based on data and assumptions. GPs allow for smooth variations in predictions, accommodating various shapes without rigid assumptions. Although advanced techniques like *t*-processes extend these ideas, the foundation largely stems from works by [24], providing a reliable understanding.

In machine learning, GPs serve as a mathematical tool for predicting real-world processes. Before training a GP model, we define its behavior using mean and covariance functions. The covariance function, or kernel, dictates how the GP learns patterns from data and makes predictions for new data points. Once the GP model is set up with training data, consisting of input-output pairs, Bayesian inference updates beliefs about the underlying process based on observed data. This yields a joint posterior distribution of function values at training and test points. Practically, these distributions are Gaussian, characterized by mean and covariance. We compute the mean and covariance of predicted function values at test points based on this joint posterior distribution. The choice of covariance function, or kernel, significantly impacts GP regression. It determines similarity between data points and influences model flexibility and generalization. Combining different kernels introduces structured patterns, enhancing the model's ability to capture complex data relationships. Therefore, to test this we introduce the Mauna Loa atmospheric $CO_2$ dataset, used by [24] to illustrate the effectiveness of a structured kernel. Fig. 1 visually depicts the time series, showcasing discernible patterns. Fig. 2 demonstrates the GP workflow, emphasizing the importance of kernel selection and hyper-parameter optimization. Each chart represents training data in green and unseen test data in blue. Our goal is to effectively model training data while accurately predicting unseen test data. The versatility of GPs lies in their kernel choices, making them a generalization of various regression techniques. Notable kernels include OU, Rational Quadratic, Matern, and Squared Exponential. For hyper-parameter optimization, conventional techniques like cross-validation or Bayesian approaches such as marginalization are employed. However, optimization may encounter non-convex likelihood functions, necessitating multiple restarts. While our focus is not on robustness under deviations from full Bayesian treatment, future research could explore this aspect further. Cross-validation, despite its challenges, offers an avenue for developing custom error functions, particularly relevant in domains like finance.

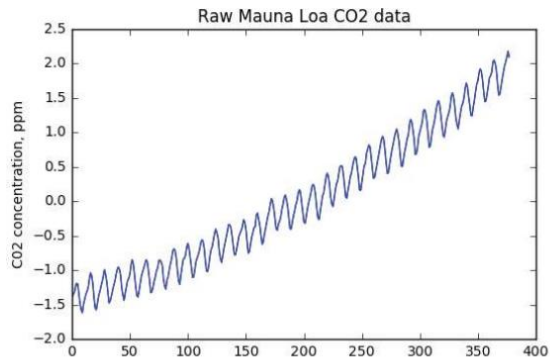

Fig. 1. Mauna Loa atmospheric $CO_2$ concentration

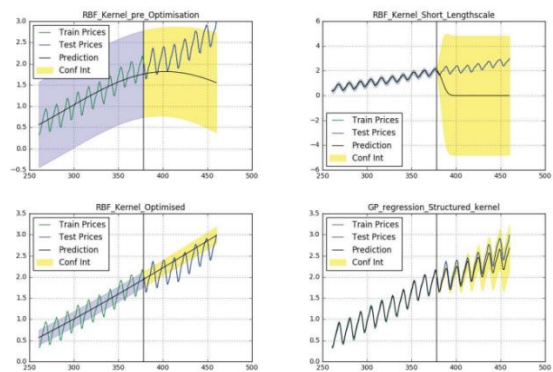

Fig. 2. The effects of length scale and kernel selection

## 3. Materials and Methods

### 3.1 Data Analysis

In this section, we discuss the ways we represent data in this study. The main idea behind these representations is to help predict a full future path. Within GPs, there's a parameter called the length-scale, which determines the distance over which data points influence each other. If the test data is much farther away from any of the training data than this length-scale, then the test data doesn't seem to relate to the existing data, and the prediction made by the GP won't be reliable. To address this issue, we adopt functional and augmented data representations, as discussed by [1] and inspired by [25]. These alternative representations aim to reduce the impact of distance on predictions and enable longer-term forecasting. We evaluate the effectiveness of these representations in Section 4. However, in our exploration of time series analysis, we have reorganized time intervals, possibly aiding the algorithm in recognizing recurring patterns over various years. Therefore, in practical scenarios such as forecasting, our dataset is likely to contain additional factors beyond just time. Following the approach outlined by [1], we represent these additional factors as $\{x\}$, where $x$ belongs to a multidimensional space. Our goal now is to determine the probability of future events, given both the explanatory variables and observed data. [1] introduces a method to incorporate these explanatory variables, which although may seem unconventional, has piqued our interest. This approach has influenced our implementation of their methodology [1] and subsequent experimental validations in Section 4. Now, considering a specific time series, denoted as $i$, and a particular time point within that series, we establish our model as per the Equations (2) and (3).

$$E[y_t^i | \mathcal{I}_{t0}^i] = f(i, t, x_{t|t_o}^i) \qquad (2)$$

$$Cov[y_t^i, y_t^{i'} | \mathcal{I}_{t0}^i] = g(i, t, x_{t|t_o}^i) \qquad (3)$$

In our approach, we consider the available information from prior historic series, denoted as series $i$ at time $t_o$. Our model is conditioned on this information. The functions $f()$ and $g()$ are derived from training our GP. We incorporate explanatory variables through $x_o$, representing a forecast of $x$ based on the information available at time $t_o$. For forecasting, we assess $f$ and $g$ with our series identity set to $N$ and time index $t$ ranging over the forecasted elements. During prediction, our information remains fixed. The augmented approach aims to include explanatory variables $x_{10}$. Since the values of $x$ are unknown during the forecasting period, we address this challenge in the subsequent section. The data representation, as elucidated by [1], relies on the observation time, which signifies when input variables are observed, conditioning our model. This time marks the last point at which explanatory variables are known. We then forecast multiple points into the future, representing the difference between the observation and target times as $A$. Augmenting the training set involves including pairs of input-output data corresponding to known explanatory variables at observation time and the target price at a future point. After training, we forecast by fixing the observation time, keeping our explanatory variables constant and allowing the target time to vary over the forecast period $t$. The model equations are then adjusted accordingly. Understanding this representation may pose some initial difficulty due to notation. We illustrate the augmented training data with simplified dummy data in Table I. Despite its simplicity, this method scales poorly for GPs due to their computational complexity. Various methods like the Nystrom method and sub-sampling have been proposed for scaling GPs. Additionally, recent advancements in String Gaussian Processes offer alternative approaches. [1] employed sub-sampling, although detailed implementation remains unspecified. We also utilized sub-sampling, facing challenges in sampling methodology, which we discuss further in the sections.

TABLE I
FUNCTIONAL AUGMENTED TRAINING DATA

| Observe Year | Observe Day | Observe Last Price | Observe Value | The Difference between two Time | Target Price |
| --- | --- | --- | --- | --- | --- |
| 7 | 5 | 787 | 15.2 | 8 | 803 |
| 7 | 5 | 787 | 15.2 | 9 | 807 |
| 7 | 5 | 787 | 15.2 | 10 | 810 |
| 7 | 6 | 783 | 15.0 | 1 | 800 |
| 7 | 6 | 783 | 15.0 | 2 | 802 |
| 7 | 6 | 783 | 15.0 | 3 | 804 |
| 7 | 6 | 783 | 15.0 | 4 | 812 |
| 7 | 6 | 783 | 15.0 | 5 | 808 |
| 7 | 6 | 783 | 15.0 | 6 | 805 |
| 7 | 6 | 783 | 15.0 | 7 | 803 |

### 3.2 Ornstein Uhelnbeck Stochastic Process

In this section, we pivot from [1] and their exploration of commodity futures, which centered on modeling real-time series with a mean reversion aspect. Our approach differs as we focus on simulated data, allowing us to manipulate variables as needed. An issue frequently encountered in time-series analysis, especially with highly noisy data, is the challenge of establishing statistical confidence in the viability of one's models. This challenge is particularly pronounced in financial time series, where models that perform well in historical back-tests often falter when applied to live data. Transitioning momentarily from machine learning to finance, it's common practice in financial modeling to employ continuous-time Stochastic Differential Equations (SDEs). For instance, the renowned Black-Scholes-Merton options pricing formula and the Vasicek interest rate model are examples of such equations. While SDEs are prevalent in finance, they lie somewhat beyond the scope of this study, except for GPs. However, to simplify matters for our purposes, let's introduce the OU process. In finance, one notable investment strategy is statistical arbitrage, wherein practitioners seek combinations of instruments that exhibit mean-reverting and stationary behavior. The premise is that significant deviations from the mean can be profitably traded, provided the parameters remain stable. The OU process serves as an example of a Brownian motion and the continuous-time equivalent of a process. This process finds applications in commodity yield modeling, such as the Gibson and Swartz model. As demonstrated in [26], an exact method exists for modeling an OU process. While a straightforward approach might involve discretizing the equation, [26] provides the exact formula for simulating an OU, which we utilize in this study. The OU model is a concept used in finance to describe how values tend to return to an average over time, even though they might move away temporarily (refer to Fig. 3). It's like a rough sketch of how things behave. When applied to financial forecasting, it's a bit pessimistic because it assumes that the only thing we know about the process is that it tends to revert to an average in a noisy way, without much structure. Imagine we're making predictions based on this model. We might do well with one set of data but not so well with another, even if the model and parameters are the same. This shows us the importance of testing against the whole range of possibilities, which a GP helps us do by offering a prediction of how the values might vary. In finance, the OU model is commonly used for mean-reverting data. We use a method called max-likelihood to fit the parameters of the model to our data, making it a straightforward and realistic way to test against GPs in later experiments. It's worth noting that the OU model requires data to be evenly spaced, which might not always be the case. However, this isn't a problem for GPs, which can handle unevenly spaced data well.

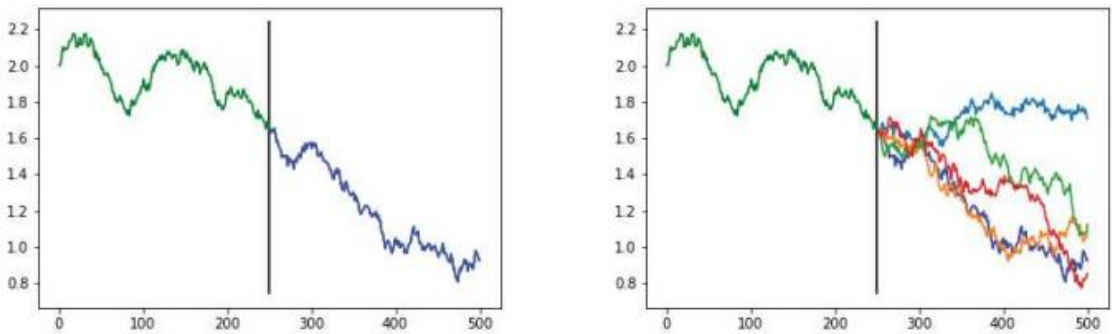

Fig. 3. OU simulation

### 3.3 Sensibility Check

In order to better understand how data functions and to ensure accuracy in our initial implementation, we analyze the Mauna Loa dataset, as utilized by [24]. Fig. 4 provides a visual representation of the data's functionality, with each year treated as an independent entity. We focus on predicting future trends from the vertical black line onwards for the final time series. In the left chart, our aim is to project the continuation of the green time series, while in the right chart, we endeavor to predict the trajectory of the blue line. Our training procedure employs a straightforward RBF GPs. Although the functional representation demonstrates a nuanced incorporation of past time series, as evidenced by the increased structural complexity compared to the simple RBF kernel illustrated in Fig. 2, there is some apparent instability. However, it's important to note that this comparison serves as an anecdotal reference rather than a statistical test.

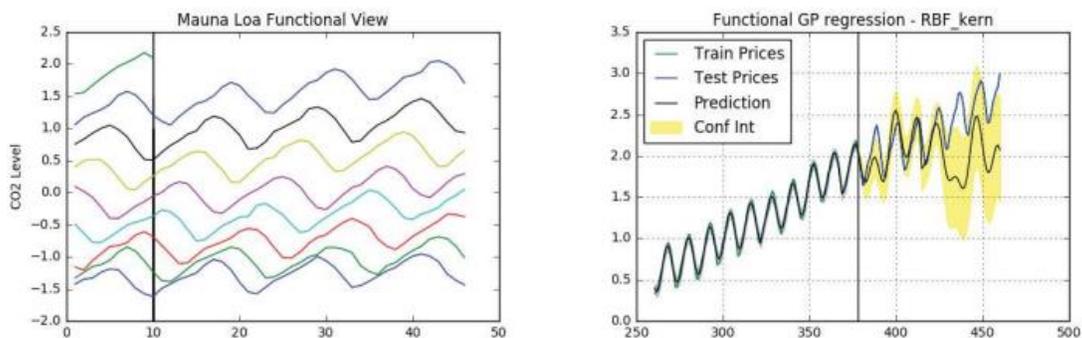

## 3.4 Sampling Analysis

The decision-making process culminated in the final choice concerning data augmentation. It revolved around three primary questions: How many samples should we include? How far into the future should we anticipate in our data augmentation? And which predictive features should we employ? In the study by [1], the sampling approach leaned towards shorter deltas (look-aheads), with regular observations made each week and multiple targets considered for each observation point. Initially, we explored a purely random sampling method, considering up to 2000 members in the training set. However, this yielded unsatisfactory outcomes. The challenge stemmed from the loss of recent information, which would have been available through our alternative data representations. For instance, if we were at year 10, day 50, attempting a forecast, the randomly sampled training set might conclude at year 9, day 247, projecting 3 days ahead. Consequently, the model would attempt predictions based on disparate contexts, leading to inaccurate forecasts. Thus, our selected sampling strategy encompassed all training set examples for each day with a zero delta, resembling conventional time series approaches. Subsequently, we augmented this data with 500 regularly sampled data points, capped with a fixed maximum delta of 50. This choice was made as we aimed to predict prices 10, 20, and 30 days into the future. However, even this pragmatic compromise imposed a significant computational burden, given the need for matrix inversions and multiple optimization restarts. The concept of delta is crucial in the augmented approach, representing the distance ahead from the observation day. It delineates the boundary beyond which the GPs lacks training examples, rendering it uninformed. Hence, the augmented approach's efficacy hinges on informative features, particularly those predictive of future target prices, such as stock valuation and momentum. For the simulated time series, features were chosen based on whether the observation point fell into one of three regimes: high and declining, low and ascending, or neither. These regimes were defined by the 50-day delta, indicating whether the series would ascend or descend over the next 50 days. Although ideal for sinusoidal time series, this approach becomes less precise with added noise, potentially yielding false signals. Inspired by real-world scenarios where signals are imperfect and noisy, the identified regimes were treated as categorical features. To facilitate interpretation within a multi-dimensional kernel, the regimes were one hot-encoded, thereby adding three features. These features were then individually treated within the kernel implementation, employing a radial basis kernel with automatic relevance determination. The remaining dimensions were subjected to kernel selection based on the specific experiment being conducted.

## 4. Experimental Analysis

### 4.1 Simulated Data

The functional data approach aims to uncover patterns over multiple years. One simple method to introduce yearly patterns is by using a repeated sine wave. Even with an inappropriate model, the functional approach might benefit from past years' data, a capability lacking in traditional one-dimensional datasets. In our simulated experiments, we introduced an OU process to add noise and simulate mean reversion. As the noise levels increased, our initially predictable seasonal time series became corrupted, dominated by the OU process. Initially, we aimed to evaluate when a standard model might outperform. However, as depicted in Fig. 5, the functional perspective showcased its advantages even with an imperfect model. We trained and optimized GPs with an OU kernel, with the last observation point represented by the black line, predicting the blue line. The GPs were trained on the green line. The top left chart illustrates a GP with an OU kernel, which did not generalize well and quickly lost effectiveness. In contrast, the top right chart, employing the same kernel but with functional data representation, yielded near-perfect predictions. The distinction between the blue test set and the black prediction was barely discernible, accompanied by a robust confidence interval. The bottom left chart presents the functional augmented view, incorporating features with a maximum delta of 50. This model performed well within the maximum delta range but declined afterward. Finally, the last chart demonstrates an OU kernel without the benefit of functional representation. Although not included in [1], we added it to distinguish between functional and augmented views for testing purposes. While it outperformed the naive one-dimensional view (top left chart) in prediction ability, the kernel was still unsuitable for this time series. Fig. 6 depicts simulated data with added OU noise ($a = 1$), significantly increasing the noise level. We applied the same GPs to this training set, as illustrated in Fig. 7, using an OU kernel on the noisy data. Although the predictions appeared similar, without proper test statistics, discerning their accuracy proves challenging.

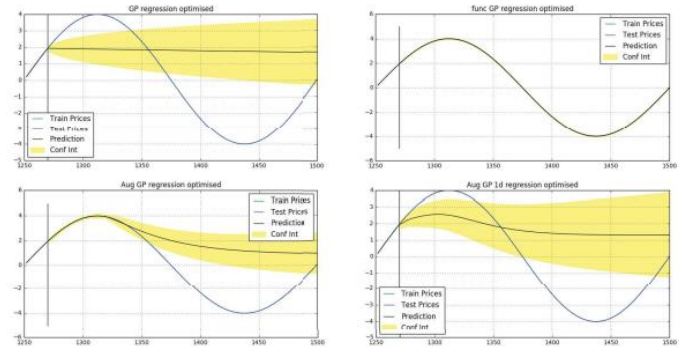

Fig. 5. GPs with OU kernel with low noise

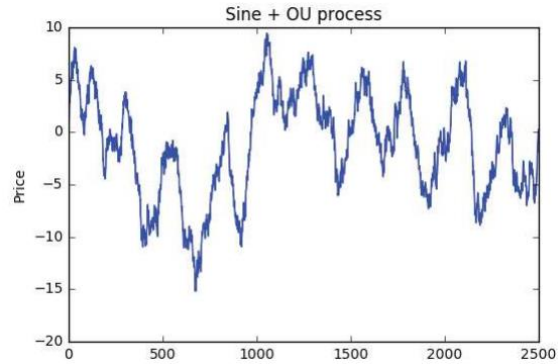

Fig. 6. Sine + OU wave

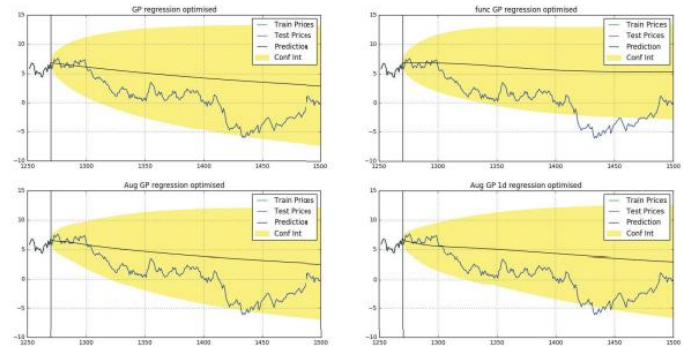

Fig. 7. GPs with OU kernel with high noise

**4.2 Test Procedure**

Tests were conducted to evaluate how well different models performed under various conditions, such as different levels of noise, different kernels, the impact of fat tails, and varying amounts of data in the training set. From our previous observations, the GPs provide a prediction of the average of our predicted distribution along with its uncertainty. Since we are simulating our time series, we can also create a test distribution. Thus, we followed the following methodology for testing:

1. We simulated a training series by generating one realization of our time series with known parameters.
2. We created training data for models, which varied based on the type of data representation: one-dimensional, functional, augmented, or augmented-functional. Additionally, predictive features were generated for the augmented approach and for sampling.
3. We trained the GPs on this training set using different restarts for each optimization.
4. We predicted the mean function and standard deviation going forwards.
5. We created a test set by simulating the test series 1000 times from the same observation point with the same parameters to create a distribution of test paths.

Fig. 8 illustrates this concept using GPs with an OU kernel to predict a noisy sine wave. The figure depicts the GPs prediction on the test set, comparing two distributions over time: the mean function for the GP (in blue) and its confidence interval (in yellow). Another comparison is made against the standard model, AR(1), calibrated via maximum likelihood. Fig. 9 presents the same models in a scenario with minimal noise. We also varied the starting levels to account for different tendencies of the model. We repeated the process 10 times to accommodate these variations. Then, we compared mean predictions at fixed points in the future (e.g., 10, 20, and 30 days) and calculated the MSE over the entire trajectory to assess prediction accuracy (Fig. 10). The functional GPs emerged as a superior predictor compared to the AR(1), even with relatively high noise levels. Additionally, we compared standard deviations of predictions to ensure proper risk management. Lastly, we varied parameters to assess their impact on the models' predictive ability, considering multiple noise levels despite the computational intensity of such analyses.

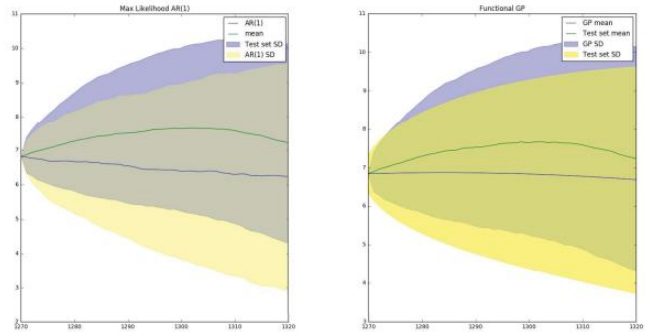

Fig. 8. AR (1) vs Functional GPs with $\sigma = 1$

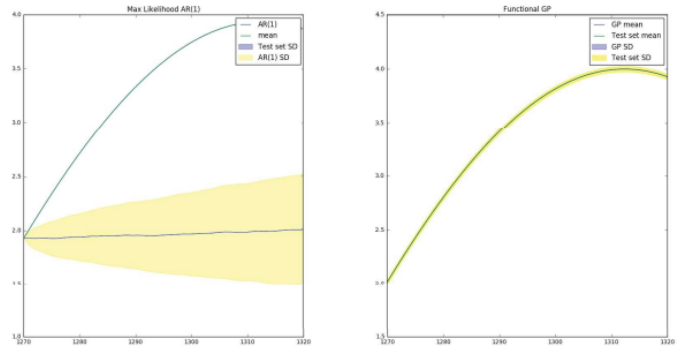

Fig. 9 AR (1) vs Functional GPs with $\sigma = 0.0001$

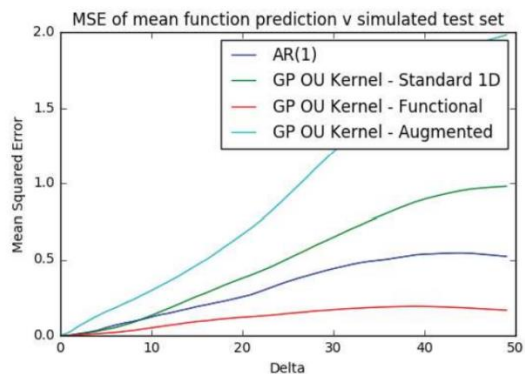

Fig. 10. MSE of mean function prediction vs simulated test set

## 4.3 Experiment 1: Noise

To understand how GPs and various data representations handle noise, researchers [1] investigated the effectiveness of an augmented-functional method applied to commodity futures using a Rational Quadratic Kernel with automatic relevance determination. The functional approach aims to predict future outcomes by utilizing past information. Augmentation, on the other hand, extends this approach by enabling predictions of multiple future time points for our features. In our experiments, we simulated [1] calendar spreads. Initially, we sought to exhibit mean reversion in our test series using an OU process. Additionally, we aimed to introduce regularities across years, achieved by combining the OU process with a sine wave. To evaluate [1] methods, we compared functional-augmentation with functional-only and augmented-only approaches. Our tests also included augmented data without the functional structure to provide clarity on the contributions of augmentation and functional methodologies (refer to Figs. 11, 12, and 13). As noise levels increase, the structure in our data deteriorates, impacting the predictive power of our models. Our experiments utilized the same kernel as [1], allowing us to compare the performance of different data representations against each other and against the commonly used AR (1) model. We observed that the functional approach outperformed the standard GPs across all noise levels, particularly excelling at low noise levels where it could predict accurately for longer durations. Similarly, the augmented representation consistently outperformed the standard GPs, raising questions about the significance of predictive features in [1] compared to the pure functional approach. Our results also indicated that the functional-augmented representation outperformed other approaches across all noise levels. However, as noise increased beyond a certain threshold, the advantages of structured views and augmented features diminished compared to the simplicity of the AR (1) model. At lower noise levels, all GPs approaches outperformed the AR (1) model significantly. Despite appearances, the AR (1) model's predictive power diminishes in deterministic processes, where GPs approaches excel. Consistency of prediction across time intervals was also observed, indicating that the best predictor over shorter durations remained the best predictor over longer durations.

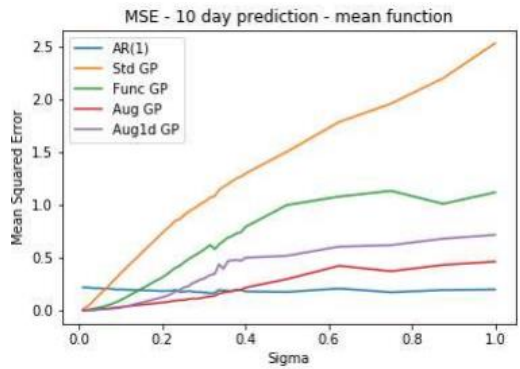

Fig. 11. MSE – 10 day prediction

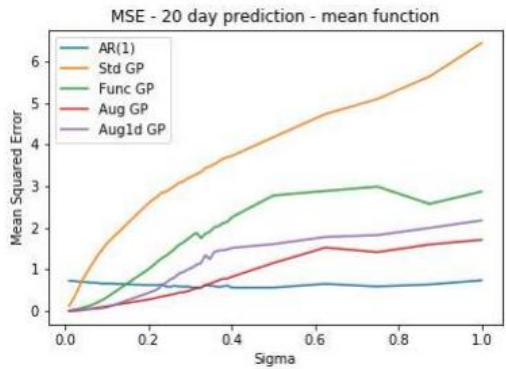

Fig. 12. MSE- 20 day prediction

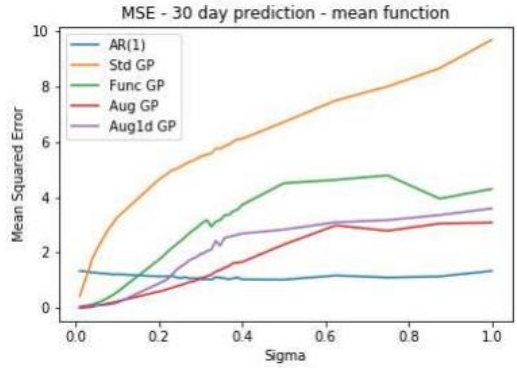

Fig. 13. MSE- 30 day prediction

### 4.4 Experiment 2: Kernel Choice

In this study, we exclusively utilize basic kernels. Our approach is inspired by previous works [1], where a rational quadratic and a squared exponential kernel were employed. While these choices are fundamental, we delve into a domain less intricate than the functional-augmented method explored in this study. Unlike methodologies leading to substantial expansions in the training set, our approach focuses on a sampling strategy essential for our study's objectives. One fundamental question arises: how critical is the selection of the kernel? To address this, we first replicate [1] experiment by employing a Rational Quadratic (RQ) kernel. Subsequently, we explore simpler alternatives, such as the Squared Exponential kernel. This study applied the functional approach to analyze stocks from Hewlett Packard, Yahoo, and Starbucks, using adjusted closing prices with an RBF kernel (Squared Exponential). Although our attempt to replicate this study faced challenges due to the lack of summary statistics in the paper, we managed to closely mirror their predictions. However, achieving this required us to artificially constrain the length-scale of the RBF kernel to incorporate the year into the functional approach. Without this constraint, the RBF GPs reverted to its prior state. Our experimentation encompassed testing the RQ, RBF, and Matern3/2 kernels, with the OU kernel representing the appropriate choice for the stochastic process we are integrating. As we explore across different kernels, we introduce a specific noise level selection, opting for $\sigma = 1$. Under this condition, the AR (1) model significantly outperformed the functional-augmented RQ kernel in our previous experiment. We aim to assess whether alternative kernels can bridge the predictive gap, especially under high noise levels. Results regarding the predictive power of various kernels unveil significant insights. The OU kernel emerges as the top performer, particularly noteworthy at high noise levels, closely matching the predictive power of an AR (1) model. Conversely, the squared exponential kernel exhibits the poorest predictive power, likely due to its excessive smoothness for our functions. Intriguingly, the performance of this inappropriate kernel substantially improved when paired with the functional-augmented structure, consistent with our findings from Experiment 1. The results with the Matern3/2 and RQ kernels align closely, with the latter, chosen by Chapados, displaying comparable predictive power to the ideal OU kernel for this process. Moreover, our findings corroborate Experiment 1, indicating enhanced predictive power across all kernels with the introduction of data structures, particularly with the functional-augmented approach. Notably, in scenarios with high noise levels and the OU kernel, the standard GP performs on par with the clever data representations, offering no additional benefits. Consistency across different time intervals further supports our conclusions, where the best predictor at 10 days remains optimal at 20 and 30 days. Tables II, III, and IV depict our simulation results and predictions, highlighting the efficacy of the OU kernel and the benefits of the functional-augmented structure.

TABLE II
MSE-SIGMA = 1 KERNEL TEST WITH MULTIPLE KERNELS -10 DAY FORECASTS

| Kernel | AR (1) | Stdanard GPs | Functional GP | Augmented GP | Augmented GP-1D |
|---|---|---|---|---|---|
| Rat-Quad | 0.200159 | 2.529615 | 1.119038 | 0.463764 | 0.716609 |
| OU | 0.241812 | 0.385954 | 0.415322 | 0.321764 | 0.575269 |
| Matern3/2 | 0.192320 | 3.173271 | 2.581655 | 0.440301 | 0.799059 |
| RBF | 0.315257 | 13.075297 | 11.401516 | 0.933333 | 2.083939 |

TABLE III
MSE-SIGMA = 1 KERNEL TEST WITH MULTIPLE KERNELS -20 DAY FORECASTS

| Kernel | AR (1) | Stdanard GPs | Functional GP | Augmented GP | Augmented GP-1D |
|---|---|---|---|---|---|
| Rat-Quad | 0.739826 | 6.439397 | 2.871008 | 1.714489 | 2.179615 |
| OU | 0.753923 | 1.298962 | 1.377844 | 1.053084 | 1.641579 |
| Matern3/2 | 0.681386 | 13.530849 | 9.950751 | 1.689285 | 2.820553 |
| RBF | 1.048223 | 34.307963 | 28.219531 | 2.985530 | 6.778340 |

TABLE IV
MSE-SIGMA = 1 KERNEL TEST WITH MULTIPLE KERNELS -30 DAY FORECASTS

| Kernel | AR (1) | Stdanard GPs | Functional GP | Augmented GP | Augmented GP-1D |
|---|---|---|---|---|---|
| Rat-Quad | 1.337654 | 9.701094 | 4.303638 | 3.086880 | 3.595598 |
| OU | Std GP | 2.389121 | 2.539594 | 1.990475 | 2.796935 |
| Matern3/2 | 9.701094 | 24.904670 | 17.563196 | 3.825518 | 6.448994 |
| RBF | 1.343196 | 43.881841 | 35.140114 | 5.501070 | 14.011281 |

**4.5 Experiment 3: Fat Tails**

In financial analysis, it's commonly observed that the behavior of market data doesn't follow a typical Gaussian pattern. Instead, it tends to have what we call 'fat tails', meaning that extreme events, like large losses, are less rare than a Gaussian distribution would predict. Our goal was to examine how these fat tails affect the performance of a specific mathematical tool called the RQ kernel, as chosen by [1]. In our first experiment, we found that when the noise level exceeded 0.4, a simple autoregressive model (AR (1)) tended to make more accurate predictions, regardless of how the data was represented. We set the noise level at a constant value of $\sigma = 0.38$. At this noise level, the RQ kernel with a special functional-augmented structure outperformed the AR (1) model as a predictor. To simulate the effect of fat-tailed noise, we employed the standard-$t$ distribution and adjusted the degrees of freedom parameter. When the degrees of freedom are high, the $t$ distribution resembles a Gaussian, but with fewer degrees of freedom, extreme events become more probable. Initially, the AR (1) model exhibited lower predictive accuracy compared to the functional-augmented GP. However, as the degrees of freedom decreased below 20, the AR (1) model demonstrated the lowest MSE in its predictions. This pattern mirrors findings from Experiment 1, where the introduction of noise led to a loss of structure, favoring the AR (1) model's predictive performance. Although the RQ kernel initially showed superior predictive ability at moderate noise levels, the AR (1) model surpassed it as the distribution's tails became heavier. While differences were observed among the standard GPs, functional GPs, augmented GPs, and functional-augmented GPs in terms of their predictive capabilities, drawing conclusive findings regarding their ability to predict the standard deviation of the test set distribution requires further analysis. When the degrees of freedom were reduced to two, all models exhibited negligible predictive ability. The resulting time series data appeared erratic, displaying significant jumps and deviating substantially from typical financial time-series patterns. In summary, introducing heavier tails into the distribution, instead of Gaussian noise, mimicked the effect of increasing volatility in the OU process, as observed in Experiment 1. Initially, the RQ-GP outperformed the AR (1) model in predicting volatility. However, as the degrees of freedom in the $t$-distribution dropped below 20, the AR (1) model demonstrated superior predictive performance. Table V present the 10-day prediction results with varying degrees of freedom.

Table V
MSE-SIGMA = 1 RQ KERNEL-10 DAY FORECASTS

| Degrees of Freedom | AR (1) | Stdanard GPs | Functional GP | Augmented GP | Augmented GP-1D |
|---|---|---|---|---|---|
| 1000 | 0.230755 | 0.979164 | 0.730086 | 0.162114 | 0.357263 |
| 100 | 0.246249 | 1.17295 | 0.782307 | 0.200303 | 0.360546 |
| 50 | 0.237387 | 0.978202 | 0.756199 | 0.249668 | 0.432981 |
| 20 | 0.271697 | 2.764549 | 0.548723 | 0.353936 | 0.381440 |
| 15 | 0.269132 | 2.91509 | 0.673308 | 0.238837 | 0.463434 |
| 5 | 0.285076 | 0.98209 | 0.575406 | 0.359739 | 0.583631 |
| 3 | 0.236719 | 1.862816 | 1.317582 | 0.548689 | 0.763959 |
| 2 | 263.41 | 32653.4 | 76371.9 | 740.3 | 48116.9 |

**5. Discussion**

The study focuses on forecasting long-term trends in time series data using GPs. Initially, it provides an overview of GPs, including their machinery and intuitive aspects. Subsequently, it introduces and implements novel data representations: functional, augmented, and functional-augmented, aiming to enhance the performance of GPs in handling large datasets. Data augmentation, a key technique, involves expanding the training set to improve GPs scalability. To validate this approach, signals are introduced to enable thorough testing of data augmentation techniques. The inspiration for this study stems from notable works such as [1]. These works highlight the potential advantages of employing GPs for financial forecasting, including Bayesian treatment, kernel usage, and hyper-parameter optimization. While the study draws from previous implementations, particularly, it deviates in its testing methodology. A novel testbed is constructed to address uncertainties inherent in forecasting time-series with stochastic elements. Instead of relying on single test realizations, the study employs simulation to generate multiple paths of the test set, enabling a comprehensive comparison with

GP predictions. In addressing real-world financial time series challenges characterized by noise, and fat tails, the study evaluates the introduced representations against varying conditions. Additionally, different kernel functions are tested to explore their impact on representation performance. Despite the importance of kernel selection, the study acknowledges the uncertainty in choosing the most suitable kernel for real-world data. Moving beyond simulated experiments, the study applies GPs-based forecasting to practical problems, such as trading on ETFs. It demonstrates the GP's ability to provide multiple forecasts along with associated uncertainties, facilitating the implementation of a trading strategy based on maximizing the expected Sharpe ratio. While the GPs exhibits promising predictive performance, it emphasizes the necessity of low execution costs for profitable trading based on this strategy.

## 6. Conclusion and Future Works

We demonstrated that our method, the functional representation, can effectively utilize past data, resulting in better predictions compared to traditional techniques. Additionally, by incorporating augmentation techniques, we observed even greater predictive accuracy, outperforming standard GPs with similar kernels. Our approach exhibited robustness against higher levels of noise and broader distribution tails. However, both representations come with trade-offs. While data augmentation enhances predictive abilities by forecasting multiple steps ahead, it also amplifies dataset size, particularly challenging for GPs due to computational complexities. The functional representation segments historical data into independent components, enabling the model to leverage past information as it progresses further away from current observations. Yet, this method imposes strong assumptions on the data, assuming significant discontinuities between consecutive time points. Regarding noise levels, at higher noise levels, structural patterns become less influential, and simpler models like AR (1) demonstrate comparable performance to GPs. Proper kernel selection is crucial; for instance, we found the RQ kernel to be robust even with substantial noise levels. Conversely, smooth kernels like the squared exponential struggled to capture underlying patterns, especially in volatile environments. Furthermore, our experiments underscored the importance of model scalability. While GPs offer rich insights, their computational demands limit their applicability to large datasets. Subsampling emerged as a workaround, albeit with its own challenges. Kernel selection remains an art, with various heuristic approaches available. Ideally, kernels should be informed by data characteristics, allowing for more structured representations. Unfortunately, this remains a complex task, although advancements such as *t*-processes show promise, particularly in domains with evolving covariance structures, like finance.